\documentclass[fleqn,usenatbib]{mnras}
\usepackage{newtxtext,newtxmath}
\usepackage[T1]{fontenc}
\usepackage{ae,aecompl}
\usepackage{graphicx}
\usepackage{amsmath}
\usepackage{siunitx}
\usepackage{xspace}
\usepackage{aas_macros}
\usepackage{grffile}
\usepackage{url}
\usepackage{xcolor}
\usepackage{hyperref}
\hypersetup{colorlinks=true, linkcolor=blue, citecolor=blue, urlcolor=blue}

\newcommand{\lm}{l_{\rm m}}
\newcommand{\msun}{\mathrm{M}_\odot}

\title[NS Magnetic Field Depth]{Correlating Surface Magnetic Field Decay with Internal Field Depth of Isolated Neutron Stars}

\author[K. Sellick \& S. Ray]{
Kathleen Sellick\thanks{E-mail: sellickk@ukzn.ac.za} and Subharthi Ray\thanks{E-mail: rays@ukzn.ac.za}
\\
Astrophysics Research Centre (ARC), Discipline of Mathematics, School of Agriculture \& Science, \\
University of KwaZulu-Natal, Durban 4001, South Africa
}

\date{Accepted 2026 September 23. Received 2026 September 22; in original form 2026 May 5}
\pubyear{2026}

\begin{document}
\label{firstpage}
\pagerange{\pageref{firstpage}--\pageref{lastpage}}
\maketitle

\begin{abstract}
The long-term evolution of neutron star crustal magnetic fields through Ohmic decay is governed by the transport properties of the crust. In this work, we demonstrate that the depth at which the magnetic flux is anchored plays a significant role in the dissipation timescale. We study the evolution of the magnetic field at the surface of the neutron star due to the crustal field evolution, under the assumption of Ohmic diffusion in a curved spacetime. We employ a cumulative distribution function to parameterise the initial current distribution and explore the effects of varying the equation of state and crustal impurity. We compare results across three neutron star models representing varying degrees of compactness. We find that magnetic fields anchored deeper within the inner crust experience negligible decay over timescales up to $10^{10}$ years. Significant field evolution, necessary to explain the deficit of high-field pulsars at advanced ages, is possible if the magnetic currents are confined to the shallow outer crust. This result holds true regardless of the impurity concentration or the specific geometric thickness of the crust determined by the equation of state.
\end{abstract}


\section{Introduction}
\label{sec:Intro}
The long-term evolution of neutron star magnetic fields remains one of the open questions in pulsar astrophysics. While the population of isolated radio pulsars exhibits a clear deficit of high magnetic field objects at advanced ages, consistent with magnetic field decay, the specific processes of this decay are debated. Although precise age estimates for individual pulsars are notoriously problematic, population synthesis studies provide broader constraints on the long-term magnetic field evolution, with several studies \citep{Gonthier_2004, 2014MNRAS.443.1891G} suggesting that, for the observed population to match the $P-\dot{P}$ diagram, the surface magnetic field must decay appreciably over a timescale of $10^6$--$10^8$ years \citep{2014MNRAS.444.1066I,2024ApJ...972...78S,2006ApJ...643..332F}.

Although magnetospheric processes are crucial to the observable emission and spin-down \citep{2017ARA&A..55..261K}, the underlying surface magnetic field evolution is governed by crustal physics and may proceed through several mechanisms operating within the star. Ohmic decay and Hall drift act within the crust, whereas ambipolar diffusion is associated with the stellar core. In more strongly magnetised neutron stars, Hall drift can become important. Although non-dissipative, it can redistribute magnetic energy from the large-scale dipolar field into toroidal and more localised magnetic structures, which can subsequently undergo more rapid Ohmic dissipation \citep{1992ApJ...395..250G}. In the crust-confined field scenario considered here, we focus on the long-term evolution of the crustal magnetic field driven by Ohmic dissipation. This process is highly sensitive to the local electrical conductivity, which rises sharply with density. Consequently, currents flowing in the highly conductive inner crust (densities $\rho \sim 10^{11}$ to $10^{14}$ g cm$^{-3}$) can persist for timescales exceeding $10^9$ years \citep{2004ApJ...609..999C}. The importance of these crustal transport properties for the long-term spin evolution is illustrated by \citet{2013NatPh...9..431P}, who used the absence of isolated X-ray pulsars with spin periods longer than $12$ s to show that sufficiently rapid magnetic field decay can reduce the spin-down torque at later ages and thereby prevent these objects from reaching substantially longer periods. In their models, this decay is enhanced by a highly resistive inner-crust layer associated with a large impurity content. In this way, they were able to establish a link between the transport properties of the inner crust of neutron stars and the timing properties of this population and provide a complementary population-level argument for magnetic field decay in highly magnetised neutron stars.

The evolution problem is further complicated by the fact that the initial configuration of the magnetic field is unknown, as its physical origin remains a subject of ongoing debate \citep{2006MNRAS.367.1323F,10.1063/1.2900262,2021Univ....7..351I}. The field may be a fossil remnant of the progenitor star, conventionally assumed to penetrate the deep interior \citep{1995JApA...16..217B}. However, as a finite interval separates the proto-neutron-star phase from the onset of type-II superconductivity, this initial flux may evolve substantially before quantised flux tubes can form \citep[e.g.,][]{2021MNRAS.503..875L}. Alternatively, the field could be confined to the crustal layers, potentially generated via thermomagnetic instabilities \citep{1983MNRAS.204.1025B, 1986MNRAS.219..703U} or early-stage convection \citep{1993ApJ...408..194T}. Conversely, if the magnetic flux remains deeply anchored within the highly conductive core, Ohmic decay is heavily suppressed, resulting in extremely long timescales \citep{1992ApJ...395..250G}. In such a case, mechanisms such as fluxoid expulsion or ambipolar diffusion may transport flux into the crust where it can then dissipate \citep{10.1046/j.1365-8711.2000.03188.x, 10.1046/j.1365-8711.2001.04469.x}. However, core composition is still uncertain and may include various superfluid, superconductive, or exotic phases. Because of this uncertainty, and because core dissipation is traditionally assumed to be slow \citep[though recent studies suggest it may proceed more rapidly depending on the internal microphysics; see e.g.,][]{2024MNRAS.528.5178S, 2026ApJ..1000..291I}, many practical models either confine the field to the crust or treat the core as effectively frozen. Furthermore, while the thermal properties of the core can influence the early stages of field evolution, the long-term Ohmic decay becomes increasingly insensitive to the cooling history as the star ages and crustal conductivity becomes dominated by density and impurity effects \citep{2012PhRvD..85j4019N}.

The dissipation of magnetic fields confined to the stellar crust has been the subject of extensive study \citep{1987ApJ...323L..61S, 1990ApJ...363..597S, 1992MNRAS.256..261U}. While these early investigations established the fundamental Ohmic processes, they were generally limited to specific stellar configurations and did not investigate how different internal structures might alter the decay. \citet{1997MNRAS.292..167U} advanced this by demonstrating that the decay is sensitive to the underlying equation of state (EoS) and the cooling scenario. While they and subsequent authors \citep{1996MNRAS.282.1059B, 2013MNRAS.434..123V} speculated that the decay rate depends appreciably on the initial depth of the currents, the physical limits of this depth and its role in reproducing observational trends were not specifically analysed. Furthermore, work by \citet{1996MNRAS.282.1059B} and \citet{1990A&A...229..133H} showed that EoS-controlled crustal thickness and impurity content strongly modulate the dissipation rate. However, the radial anchoring depth has rarely been treated as the primary independent variable in a systematic survey across a wide range of modern stellar configurations. Our study builds upon these foundations by providing a dedicated, systematic survey of the radial anchoring depth ($\lm$) within a curved spacetime framework. We test these geometric requirements across a comprehensive range of stellar structures obtained from two known EoSs \citep{1974AnPhy..83..491W,1974NuPhA.230....1B}, and an extreme, observationally-constrained reconstructed EoS for a high-mass model \citep{2024PhRvD.109d3052F}. This allows us to determine if the requirement for a shallow magnetic field is consistent across different regimes of gravitational curvature and crustal properties.

We calculate the magnetic field evolution using an initial profile based on a cumulative distribution function (CDF). By evaluating the decay across varying anchoring depths ($\lm$) and impurity parameters ($Q$) across our range of stellar models, we find that the results are highly sensitive to the initial current distribution. Our results show that regardless of the specific parameters, physically realistic decay rates are possible if the magnetic flux is confined to the shallow outer crust.

This article is organised as follows. In Section 2, we present the basic formalism, defining the general relativistic induction equation, the parameterisation of the initial magnetic field configuration, and the microphysical transport properties governing the crustal conductivity. Section 3 details the physical models and equations of state used to determine the mass density profiles and resulting crustal structures. In Section 4, we present the results, exploring how the magnetic field depth and crustal impurities dictate the timescale of Ohmic dissipation across varying degrees of stellar compactness. Finally, in Section 5, we discuss the astrophysical implications of our results for the long-term evolution of isolated neutron star populations.

\section{Basic Formalism}
\label{sec:BF}
In isolated pulsars, the magnetic flux slowly diffuses through the highly degenerate stellar material, dissipating energy via Ohmic decay. To accurately model this temporal evolution, it is essential to construct a framework that couples the depth-dependent transport properties of the dense matter with the electromagnetic equations under a strong gravitational field consideration. In this section, we detail the physical model governing the crustal field decay. We begin by defining the general relativistic induction equation, outline the parameterisation of the initial magnetic field depth, and formally describe the electrical conductivity profile that fundamentally regulates the dissipation timescale.

\subsection{Magnetic field evolution}
\label{subsec:Evolution}
To describe the Ohmic evolution of the crustal magnetic field in curved
spacetime, we derive the corresponding induction equation using the covariant
form of Maxwell's equations:
\begin{equation}
\label{eq:covMax1}
\frac{1}{\sqrt{-g}}
\frac{\partial}{\partial x^\nu}
\left(\sqrt{-g}F^{\mu\nu}\right)
=
-\frac{4\pi}{c}J^\mu
\end{equation}
and
\begin{equation}
\label{eq:covMax2}
\frac{\partial F^{\mu\nu}}{\partial x^\lambda}
+
\frac{\partial F^{\nu\lambda}}{\partial x^\mu}
+
\frac{\partial F^{\lambda\mu}}{\partial x^\nu}
=0.
\end{equation}

The generalised Ohm's law is given by:
\begin{equation}
\label{eq:genOhm}
J^\mu
=
\sigma g^{\mu\nu}F_{\nu\lambda}u^\lambda,
\end{equation}
where $F_{\mu\nu}$ are the components of the electromagnetic field tensor,
$J^\mu$ are the components of the four-current density, $u^\mu$ are the
components of the fluid four-velocity, $g_{\mu\nu}$ are the components of the
spacetime metric describing the background geometry, and
$g=\det\lvert g_{\mu\nu} \rvert$. Here, Greek indices denote four-dimensional spacetime
coordinates, while Latin indices denote the three spatial coordinates.

Assuming a stationary gravitational field and negligible hydrodynamic motion within the crust (taking $u^i=dx^i/ds=0$), the corresponding induction equation in curved spacetime can be derived using equations~\eqref{eq:covMax1}--\eqref{eq:genOhm} as \citep{1997ApJ...479L.133S,2024MNRAS.528.3163S}:
\begin{flalign}
\label{eq:indcurve}
\frac{\partial F_{kj}}{c\partial t}
&=
\frac{c}{4\pi}
\left\{
\frac{\partial}{\partial x^k}
\left[
\frac{1}{\sqrt{-g}}
\frac{1}{\sigma u^0}
g_{ij}
\frac{\partial}{\partial x^l}
\left(\sqrt{-g}F^{il}\right)
\right]
\right.
\nonumber\\
&\qquad\left.
-
\frac{\partial}{\partial x^j}
\left[
\frac{1}{\sqrt{-g}}
\frac{1}{\sigma u^0}
g_{ik}
\frac{\partial}{\partial x^l}
\left(\sqrt{-g}F^{il}\right)
\right]
\right\}.
&
\end{flalign}

The magnetic field strengths considered in this work
($B\sim10^{10}$--$10^{12}$~G) are sufficiently weak that their contribution to the background spacetime curvature may be neglected. Furthermore, the crust
contains only a small fraction of the total gravitational mass, such
that its self-gravity is negligible compared with that generated by the
stellar interior. We therefore describe the crustal region by a static,
spherically symmetric Schwarzschild background, whose line element is:
\begin{equation}
\label{eq:metric}
ds^2
=
\left(1-\frac{2m}{r}\right)c^2dt^2
-
\left(1-\frac{2m}{r}\right)^{-1}dr^2
-
r^2\left(d\theta^2+\sin^2\theta\,d\phi^2\right)
\end{equation}
where $m=GM/c^2$, with $M$ denoting the total gravitational
mass of the core. Since the crust makes up only a few percent of the total gravitational mass, $M$ can be considered as the total stellar mass. For matter at rest in these coordinates, the Schwarzschild
metric gives $u^0=\left(1-\frac{2m}{r}\right)^{-1/2}$.

The electromagnetic field expressed in the Schwarzschild coordinate basis may be related to the field measured in a locally inertial Lorentz frame through an orthonormal tetrad. If $F_{(\alpha\beta)}$ denotes the components of the electromagnetic field tensor in this local frame, then
\begin{equation}
\label{eq:tetrad_transform}
F_{(\alpha\beta)}
=
e^\gamma{}_{(\alpha)}
e^\delta{}_{(\beta)}
F_{\gamma\delta},
\end{equation}
where $e^\gamma{}_{(\alpha)}$ are the components of the orthonormal tetrad associated with the Schwarzschild geometry, and parenthesised indices denote components measured in the local Lorentz frame. The non-zero tetrad components are
\[
e^t{}_{(t)}
=
\left(1-\frac{2m}{r}\right)^{-1/2},
\qquad
e^r{}_{(r)}
=
\left(1-\frac{2m}{r}\right)^{1/2},
\]
\[
e^\theta{}_{(\theta)}=\frac{1}{r},
\qquad
e^\phi{}_{(\phi)}=\frac{1}{r\sin\theta}.
\]

We now restrict the magnetic configuration to an axisymmetric dipole. The vector potential may therefore be written in spherical polar coordinates as $\mathbf{A}=(0,0,A_\phi)$, with $A_\phi=A_\phi(r,\theta,t)$. Using $F_{\mu\nu}=A_{\nu,\mu}-A_{\mu,\nu}$ and substituting the Schwarzschild metric into equation~\eqref{eq:indcurve}, the induction equation can be expressed directly in terms of the azimuthal vector potential as \citep{1997ApJ...479L.133S,2024MNRAS.528.3163S}:
\begin{equation}
\label{eq:Aphi_induction}
\begin{aligned}
\frac{\partial A_\phi}{\partial t}
&=
\frac{c^2}{4\pi\sigma}
\left(1-\frac{2m}{r}\right)^{1/2}
\sin\theta
\left\{
\frac{\partial}{\partial r}
\left[
\left(1-\frac{2m}{r}\right)
\frac{1}{\sin\theta}
\frac{\partial A_\phi}{\partial r}
\right]
\right.
\\
&\qquad\left.
+
\frac{\partial}{\partial\theta}
\left[
\frac{1}{r^2\sin\theta}
\frac{\partial A_\phi}{\partial\theta}
\right]
\right\}.
\end{aligned}
\end{equation}

For a dipolar magnetic field in Schwarzschild geometry, we choose the azimuthal vector potential in the form $A_\phi(r,\theta,t)=-f(r,t)\sin^2\theta$. Substituting this into equation~\eqref{eq:Aphi_induction}, and introducing the dimensionless radial coordinate $x=r/R$ and the compactness parameter $y=2m/R=2GM/(Rc^2)$, gives
\begin{equation}
\label{eq:diffusion}
\begin{aligned}
\frac{\partial f(x,t)}{\partial t}
&=
\frac{c^2}{4\pi R^2\sigma(x)}
\left(1-\frac{y}{x}\right)^{1/2}
\left[
\left(1-\frac{y}{x}\right)
\frac{\partial^2f(x,t)}{\partial x^2}
\right.
\\
&\qquad\left.
+
\frac{y}{x^2}
\frac{\partial f(x,t)}{\partial x}
-
\frac{2}{x^2}f(x,t)
\right].
\end{aligned}
\end{equation}
where $R$ is the stellar radius and $\sigma(x)$ is the depth-dependent electrical conductivity profile of the crust, described in Section~\ref{subsec:conductivity}.

The boundary conditions for this evolution are given by $R\partial f/\partial r+f=0$ at the surface ($r=R$), which physically corresponds to matching the internal field to a current-free vacuum dipole exterior. The inner boundary condition is $f=0$ at the crust bottom, implying that the magnetic flux is strictly confined to the crust and does not penetrate the highly conductive core.

\vskip .5cm

\subsection{Initial Magnetic Field Configuration}
\label{subsec:IFP}
To determine the effect of the magnetic field depth on the diffusion rate, we construct an initial field profile that is smoothly confined to the crust. The initial configuration is derived from the standard stationary solution for a dipole field in Schwarzschild geometry, modulated by a radial distribution of the field using a cumulative distribution function (CDF) profile \citep{1996MNRAS.282.1059B}. This specific distribution provides a mathematically smooth and highly controlled method to systematically vary the depth of the field. The initial profile $f(x,0)$ is given by:
\begin{equation}
\label{eq:IP}
f(x,0) = \frac{3R^3x f_{\scriptscriptstyle \mathrm{CDF}}(x)}{8m^3} \left[ x^2\ln{\left( 1 - \frac{2m}{Rx} \right)} + \frac{2m}{R}x + 2\left(\frac{m}{R}\right)^2 \right],
\end{equation}
with the CDF given by:
\begin{equation}
\label{eq:CDF}
f_{\scriptscriptstyle \mathrm{CDF}}(x,0) = \mathcal{N} \;\Phi\!\left( \frac{x - x_{\rm turn}}{x_{\rm w}} \right).
\end{equation}

In this expression, $\mathcal{N}$ is a normalisation constant, and $\Phi$ is the CDF of the standard normal distribution. The radial turning point $x_{\rm turn} = 1 + \lm(x_{\rm min} - 1)$ defines the depth at which the magnetic current distribution is centred, while $x_{\rm w}$ characterises the width of the current layer. This CDF allows for the physical confinement of magnetic currents to specific radial depths, with $\lm$ serving as the primary control parameter: larger values of $\lm$ place the field peak deeper in the highly conductive inner crust, while smaller values confine it to the resistive outer layers. An illustrative example of the resulting initial distribution function $f(x,0)$, normalised to its surface value, is presented for different choices of $\lm$ in Section~\ref{sec:results}.

\subsection{Crustal Conductivity and Structure}
\label{subsec:conductivity}
The decay rate of the magnetic field is regulated by the electrical conductivity profile $\sigma(x)$. Because the magnetic field evolution is highly sensitive to the depth-dependent transport properties, the local conductivity is directly dictated by the mass density profile $\rho(x)$ of the neutron star (derived subsequently in Section \ref{sec:density_model}). We model the crust as two distinct regions: the outer and inner crust.

\subsubsection{Outer Crust}
\label{subsubsec:Outer}
In the outer crust ($x > x_{\rm drip}$, where $x_{\rm drip}$ is the transition point from the outer to inner crust at the neutron drip density), conductivity is governed by electron-phonon scattering ($\sigma_{\rm ph}$) and electron-impurity scattering ($\sigma_{\rm imp}$). The net outer conductivity is computed using:
\begin{equation}
\sigma_{\rm outer} = \left(\frac{1}{\sigma_{\rm ph}} + \frac{1}{\sigma_{\rm imp}} \right)^{-1}.
\end{equation}

Both scattering components are strongly dependent on the local density $\rho(x)$, which is parameterised as \citep{1979Afz....15..647U,1992MNRAS.256..261U}:
\begin{equation}
\label{eq:chi}
\chi = \left( \frac{\rho(x)}{\mu_e 10^6 \text{ g cm}^{-3}} \right)^{\frac{1}{3}},
\end{equation}
where $\mu_e = A/Z$ is the mean molecular weight per electron, with $A$ and $Z$ being the mass number and atomic number
respectively. For simplicity, we assume that only one species of ions is present.

The phonon scattering component \citep{1980AZh....57..526I} is defined as:
\begin{equation}
\sigma_{\rm ph} \approx \frac{1.57 \times 10^{23} \chi^4}{T_6(2 + \chi^2)} \frac{\sqrt{0.017 + \delta^2}}{13\delta} ~~\text{ s}^{-1},
\end{equation}
where $T_6 = T/(10^6\,{\rm K})$ and $\delta = 0.45(T/T_D)$. The Debye temperature $T_D$ also scales continuously with the density parameter $\chi$:
\begin{equation}
\label{eq:Debye}
T_D = 0.45 \frac{\hbar}{k_B}\left( \frac{4\pi Z^2 e^2 n_i}{A m_p} \right)^{\frac{1}{2}} = 2.4 \times 10^6 \left( \frac{2}{\mu_e} \right)^{\frac{1}{2}} \chi^{\frac{3}{2}} \text{ K},
\end{equation}
where $k_B$ is the Boltzmann constant, $m_p$ is the proton mass and $n_i$ is the ion number density. 

At lower temperatures and higher densities, the impurity scattering component dominates and is defined as:
\begin{equation}
\sigma_{\rm imp} = \frac{8.53 \times 10^{21} \chi^3}{\Lambda_{\rm imp} \left( 1 + \chi^2 \right)} \frac{Z}{Q} ~~\text{ s}^{-1},
\end{equation}
where $Q$ is the impurity parameter and $\Lambda_{\rm imp}$ is the Coulomb logarithm, which is $\Lambda_{\rm imp} \cong 2$ for $\rho \geq 10^5 ~~\text{ g cm}^{-3}$.

Although a neutron star is born at extreme temperatures ($T \sim 10^{9}$--$10^{10}$ K) \citep{2004ARA&A..42..169Y}, it cools rapidly via neutrino emission, primarily through URCA processes in the core and neutrino bremsstrahlung in the crust. Previous studies incorporating rotational deformation and thermal evolution \citep{2012PhRvD..85j4019N} indicate that by $t \sim 10^2$ yr, the internal temperature stabilises around $10^{5.2}$ K. Long-term cooling curves \citep{2006NuPhA.777..497P, 2013MNRAS.434..123V} further demonstrate that temperatures remain near $10^5$ K for the vast majority of the pulsar's active lifetime ($10^6$--$10^{8}$ yr). Furthermore, while some models suggest that the polar caps may experience localised heating due to the continuous bombardment of back-flowing antiparticles from the magnetosphere \citep{2017AstL...43..820T, 2020MNRAS.493.3770S}, this ongoing process still allows the broader crustal system to establish and maintain a steady thermal equilibrium over the vast majority of its lifespan.

Because isolated radio pulsars lack external heating sources and rapidly settle into this quasi-equilibrium thermal state, the long-term Ohmic dissipation is effectively governed by a quasi-steady temperature. Therefore, we adopt an isothermal approximation for the crust, assuming a temperature of $T = 10^{5.5}$ K, accommodating for the effects of the back-flowing particles at the polar caps. The validity of this isothermal assumption is supported by the disparity in relevant timescales: the crustal thermal relaxation timescale, typically $\tau_{\rm th} \sim 10^2$--$10^3$ years \citep{2013MNRAS.434..123V}, is negligible compared to the magnetic diffusion timescales ($\sim 10^6$--$10^9$ years) considered in this work. Consequently, localised thermal energy deposited at the polar caps is efficiently redistributed throughout the entire crust \citep[see, e.g.,][]{2020MNRAS.493.3770S}. This rapid global redistribution slightly elevates the baseline thermal equilibrium, physically justifying our use of a uniform $10^{5.5}$ K profile rather than the estimated cooling limit of $\sim 10^{5.2}$ K as given in \cite{2012PhRvD..85j4019N}.

\subsubsection{Inner Crust}
\label{subsubsec:Inner}
As density increases deeper within the neutron star crust, the ground-state ratio $Z/A$ decreases and neutrons become progressively less bound. Because the resulting neutron \emph{gas} cannot be reproduced in terrestrial laboratories, the physical modelling of the inner crust ($x \leq x_{\rm drip}$) must be modelled theoretically. While traditional calculations rely on classical liquid drop or semi-classical Thomas--Fermi approximations, more robust studies employ self-consistent mean-field methods, which are highly regarded for their precision in describing heavy laboratory nuclei \citep{2008LRR....11...10C}. This inner crust region is dominated by the neutron drip regime, characterised by the presence of unbound neutrons that may become superfluid. We model this region by adopting the number-density relation prescribed by \citet{GourgouliatosCumming2014}. This provides a computationally feasible approximation for the transport properties of the ultra-relativistic, degenerate electron gas, where conductivity scales with the electron Fermi momentum ($p_F \propto n_e^{1/3}$). Specifically, the electrical conductivity is scaled continuously from the drip boundary value, $\sigma_{\rm drip}$, assuming a dependence on the electron number density $n_e \propto \rho$:
\begin{equation}
\sigma_{\rm inner} = \sigma_{\rm drip} \left( \frac{\rho(x)}{\rho_{\rm drip}} \right)^{2/3}.
\end{equation}

To prevent numerical instabilities, the inner and outer conductivity regimes are smoothly blended across the neutron drip boundary.

\section{Physical Model and Equation of State}
\label{sec:density_model}
The mass density profile $\rho(x)$ required to calculate the electrical conductivity is determined by the equation of state (EoS) of dense matter. Unlike the deep stellar core, the EoS of crustal matter is better understood and more widely accepted, as the relevant nucleon interactions are known and robust many-body theory techniques are available \citep{2008LRR....11...10C}. 

Because of the assumption that the magnetic field is restricted to the crust, we only require the resulting Mass-Radius ($M-R$) parameters obtained from the different nuclear matter EoSs. We combine the standard crustal density profile with three separate stellar models each governed by a nuclear matter EoS representing different regimes of stiffness. These resulting $M-R$ relations dictate the stellar physical parameters, thereby defining the overall size of the star and the thickness of the crust. For our work, we chose the $M-R$ values from three EoS models, which are the Walecka (WAL) EoS \citep{1974AnPhy..83..491W}, the Bethe-Johnson V (BJV) EoS \citep{1974NuPhA.230....1B}, and a reconstructed EoS from a higher mass star (HMS) model as determined in \citet{2024PhRvD.109d3052F}.

We make use of the analytical formulae provided by the SLy4 EoS model \cite{2001A&A...380..151D, 2004A&A...428..191H}. While the SLy4 model provides an EoS for the entire star, we utilise its analytical formulation strictly to find the pressure-density relation within the crust. Letting $\zeta = \log_{10}(\rho)$, the smoothed pressure fit $P_{\rm fit}$ is expressed as:
\begin{equation}
    \log_{10} P_{\rm fit} = \sum_{i=1}^{4} p_{0i}(\zeta) s_i(\zeta),
\end{equation}
where $p_{0i}(\zeta)$ are polynomial coefficients specific to the SLy4 parameterisation, and $s_i(\zeta)$ are smoothing functions defined by:
\begin{equation}
    s_i(\zeta) = \frac{1}{\exp[a_i(\zeta - b_i)] + 1},
\end{equation}
with $a_i$ and $b_i$ functioning as fitting constants. To extend the physical validity to the low-density outer envelope, this analytical fit is mathematically matched to a temperature-dependent pressure approximation for an iron envelope \citep{2004A&A...428..191H}. The total pressure $P$ as given by Sly4 EoS is:
\begin{equation}
    P = P_{\rm fit} + 3.5 \times 10^{14} \rho.
\end{equation}
We use this relation from the surface to the bottom of the crust where the density reaches the normal nuclear matter density.

With the global mass $M$ and radius $R$ established by our three chosen NS models (WAL, BJV, and HMS), and the $P(\rho)$ relation defined by the SLy4 crustal formula, we obtain the physical density profile of the crust. Rather than integrating the full Tolman-Oppenheimer-Volkoff (TOV) equations from the stellar centre, we integrate a simplified hydrostatic equilibrium equation from the surface inwards. This approach is highly accurate for the crustal domain because the mass contained within the crust is negligible compared to the total stellar mass ($m(r) \approx M$), and the local pressure is negligible compared to the rest-mass energy density ($P \ll \rho c^2$). Consequently, the pressure contributions to the gravitational field vanish, and the TOV equation reduces to:

\begin{equation}
    dz = \frac{dP}{\bar{\rho} g_{\rm local}},
\end{equation}

where $z = R - r$ is the depth from the surface, $\bar{\rho}$ is the local geometric mean of the density between integration steps, and the local gravitational acceleration incorporating the general relativistic redshift is given by:

\begin{equation}
    g_{\rm local} = \frac{GM}{r^2} \left( 1 - \frac{2GM}{rc^2} \right)^{-1/2}.
\end{equation}

This procedure allows us to map the physical density $\rho$ to the normalised radial coordinate $x = 1 - (z/R)$ for each of the three EoS models. The varying $(M-R)$ parameters of the WAL, BJV, and HMS models naturally dictate the different resulting crustal thicknesses, and allow us to accurately locate the structural transition at neutron drip ($x_{\rm drip}$) where the integrated density matches $\rho_{\rm drip}$.

\section{Results}
\label{sec:results}
We solve the diffusion equation \eqref{eq:diffusion} using the \emph{method of lines} \citep{10.5555/1403886}. The spatial derivatives are discretised using finite differences on a uniform radial grid spanning the crustal domain. The resulting system of ordinary differential equations is integrated in time using a stiff-capable solver. 

We vary the impurity parameter across values of $Q \in [0.001, 0.01, 0.1]$ along with variations of the field depth parameter $\lm \in [0.001, 0.1, 0.3, 0.5, 0.8]$. This is done for the three distinct EoS models representing different regimes of stellar mass and stiffness.

The Walecka and Bethe-Johnson V models serve as our standard configurations, derived from nuclear interactions to yield canonical $1.4 \, \msun$ stars. To ensure our magnetic field evolution trends remain consistent across extreme structural regimes, we contrast these with the HMS model. The HMS configuration by \citet{2024PhRvD.109d3052F} utilises data-driven reconstruction by using multi-messenger observational data, including gravitational wave events and precise pulsar mass-radius measurements, alongside high-energy theoretical constraints where they provide an observationally grounded maximum-mass configuration.

For all models, the crustal boundaries are determined by density thresholds. We define the transition from the outer to the inner crust at $x_{\rm drip}$ where $\rho = 4 \times 10^{11} \text{ g cm}^{-3}$ (approximate neutron drip density), and the bottom of the crust (crust-core interface) at $x_{\rm min}$ where $\rho = 1.5 \times 10^{14} \text{ g cm}^{-3}$ (approximate normal nuclear matter density). The crustal characteristics of each model are given as follows:

\begin{itemize}
    \item \emph{Walecka Model:} A relativistic mean-field model from where we have used the canonical mass of $1.4 \, \msun$ corresponding to a radius of $R = 12.28$ km. In this configuration, the transition from outer to inner crust occurs at a normalised radius $x_{\rm drip} \approx 0.956$ and the crust extends down to the crust-core interface at $x_{\rm min} \approx 0.892$. This corresponds to a full physical crust thickness (outer + inner) of $\Delta R \approx 1.33$ km.
    
    \item \emph{Bethe-Johnson V:} An intermediate EoS for which we also use a mass of $1.4 \, \msun$ but which yields a more compact radius of $R = 10.691$ km. The outer-inner crust transition occurs at $x_{\rm drip} \approx 0.963$, and the crust-core interface is located deeper relative to the total radius at $x_{\rm min} \approx 0.908$. This results in a physical crust thickness of $\Delta R \approx 0.98$ km.
    
    \item \emph{Higher Mass Star Model:} An observationally constrained EoS configuration resulting in a massive star of $2.25 \, \msun$ with a radius of $R = 11.9$ km \citep{2024PhRvD.109d3052F}. Due to the intense gravitational compression at this high mass, the crust is physically thinner. The outer-inner crust transition occurs at $x_{\rm drip} \approx 0.978$, and the crust-core interface is located at $x_{\rm min} \approx 0.944$. This results in a significantly reduced physical crust thickness of $\Delta R \approx 0.67$ km.
\end{itemize}

We compare the magnetic field evolution across these varying degrees of crustal compression and thickness, from the extended crust of the Walecka model, to the intermediate Bethe-Johnson V, and the highly gravitationally compressed HMS crust, isolating the role of relative depth ($\lm$) against physical variations.

Our results are taken in terms of the normalised surface magnetic field strength. Because the Ohmic decay equation is linear with respect to $\mathbf{B}$ (in the absence of Hall drift), the initial magnetic field strength $B_0$ acts merely as a scaling factor. As such, it becomes a redundant exercise to examine multiple absolute initial field strengths. Furthermore, although the different initial radial configurations do not contain exactly the same total magnetic energy, rescaling them to a common initial energy would only change their absolute field amplitudes and would leave the normalised evolution $B_{\rm surf}(t)/B_{\rm surf}(0)$, and hence the inferred decay timescales, unchanged. We therefore use a single value of $B_0$ for all configurations.

\begin{figure}
    \centering
    \includegraphics[width=\columnwidth]{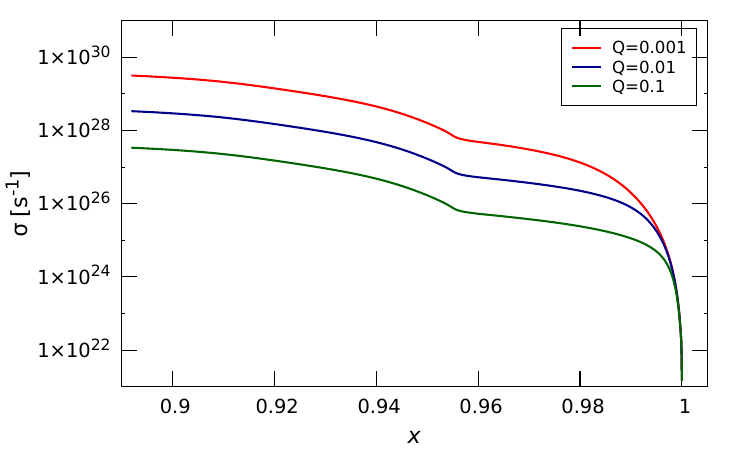} \\
    \includegraphics[width=\columnwidth]{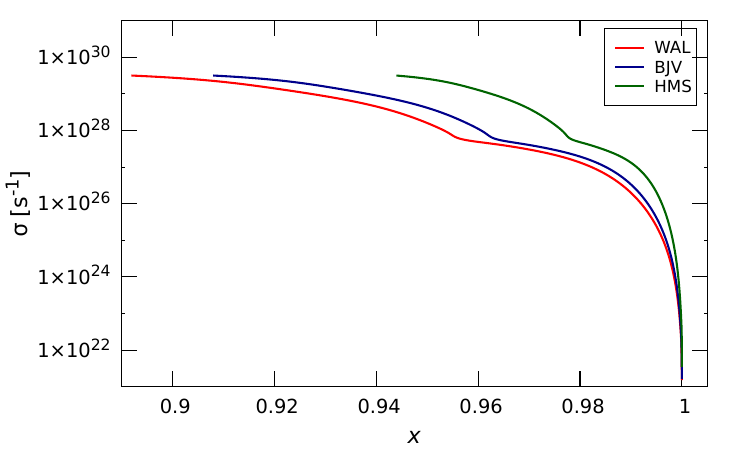}
    \caption{Electrical conductivity through the crust. Top: The Walecka EoS for varying impurity parameters $Q \in [0.1, 0.01, 0.001]$. Bottom: The electrical conductivity for $Q=0.001$ across all three EoS models.}
    \label{fig:sigma}
\end{figure}

We also examine the transport properties that dictate the Ohmic diffusion rate. The electrical conductivity through the crust is shown in Figure~\ref{fig:sigma}. The top panel illustrates the effect of the impurity parameter $Q$ on the Walecka model to show the sensitivity of the conductivity to $Q$. A highly impure crust ($Q=0.1$) possesses an electrical conductivity roughly $1-2$ orders of magnitude lower than a purer crust ($Q=0.001$). 

The bottom panel compares the $\sigma(x)$ profiles across the three stellar models for a fixed $Q=0.001$. While the absolute magnitudes of the inner and outer conductivities are similar, the distribution is more compressed for more compact models. 

\begin{figure}
    \centering
    \includegraphics[width=\columnwidth]{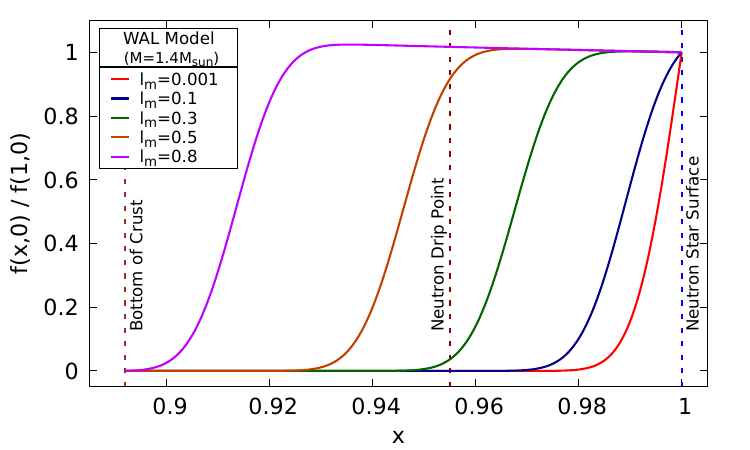} 
    \caption{The initial distribution function $f(x,0)$, normalised to its surface value $f(1,0)$, plotted against the normalised radius for the Walecka EoS model. This illustrates how varying the depth parameter $\lm$ systematically shifts the radial placement of the profile within the crust.}
    \label{fig:fx_distribution}
\end{figure}

\begin{figure}
    \centering
    \includegraphics[width=\columnwidth]{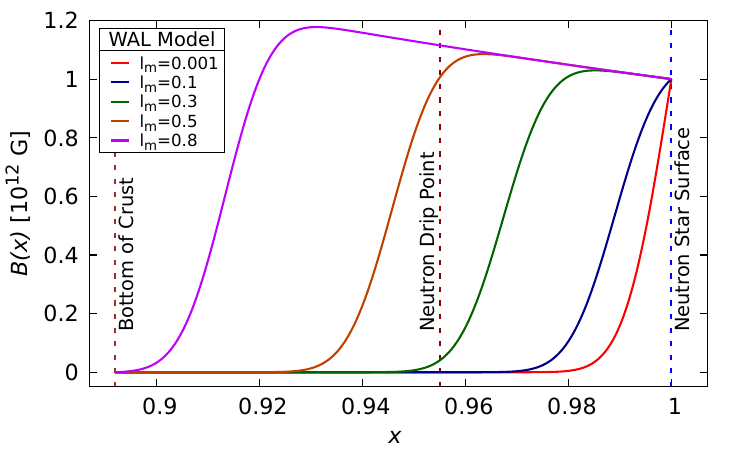} \\
    \includegraphics[width=\columnwidth]{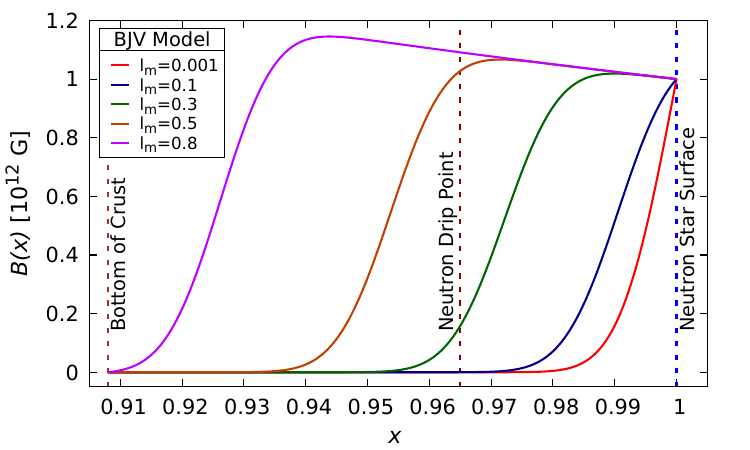} \\
    \includegraphics[width=\columnwidth]{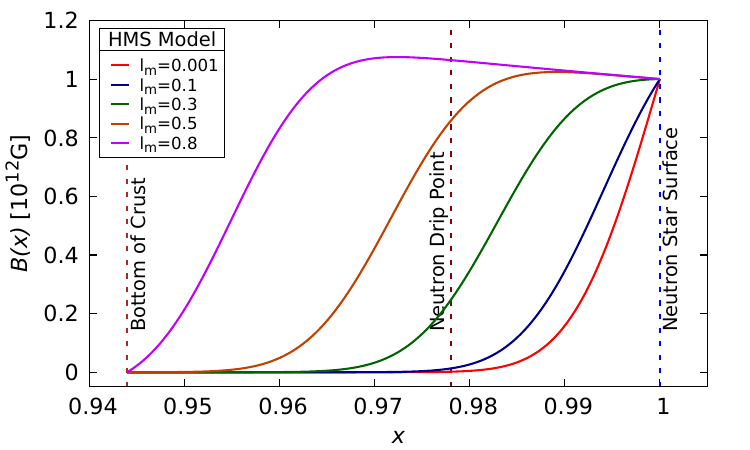} \\
    \caption{Initial magnetic field profiles for the Walecka (top), Bethe-Johnson V (middle), and HMS (bottom) models for different values of the depth parameter $\lm$.}
    \label{fig:initial_profiles}
\end{figure}

The initial magnetic field profile is defined, as in equation \eqref{eq:CDF}, using a depth parameter $\lm$, which dictates the turning point of the function within the crust. To explicitly illustrate the behaviour of this underlying stream function, Figure~\ref{fig:fx_distribution} presents the initial distribution $f(x,0)$ for the Walecka EoS across various choices of $\lm$, normalised to its surface value, $f(x,0)/f(1,0)$. The figure therefore illustrates the radial structure and relative placement of the profiles for different values of $\lm$. Small $\lm$ values correspond to currents more confined to the less conductive surface, while large $\lm$ values place the field deeper in the crust. Because the radial magnetic field is directly proportional to this stream function (e.g., $B_r \propto f(x)/x^2$), the physical magnetic profile closely resembles this S-shaped geometry. Figure~\ref{fig:initial_profiles} displays these resulting initial magnetic configurations across the three EoS models, with dashed vertical lines denoting the structural boundaries of the crust.

\begin{figure}
    \centering
    \includegraphics[width=\columnwidth]{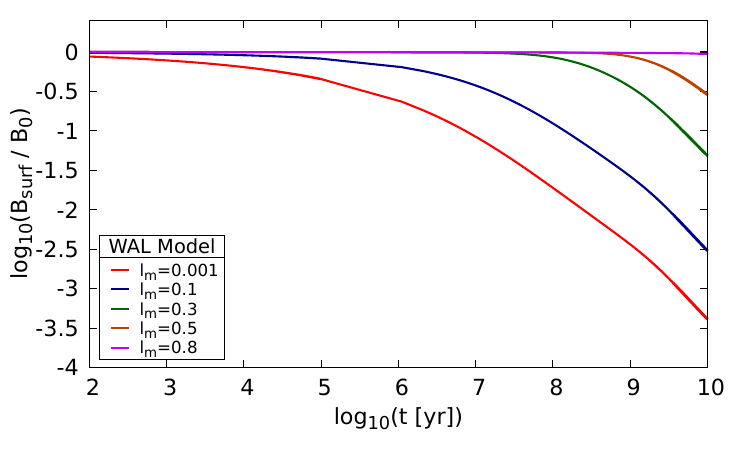} \\
    \includegraphics[width=\columnwidth]{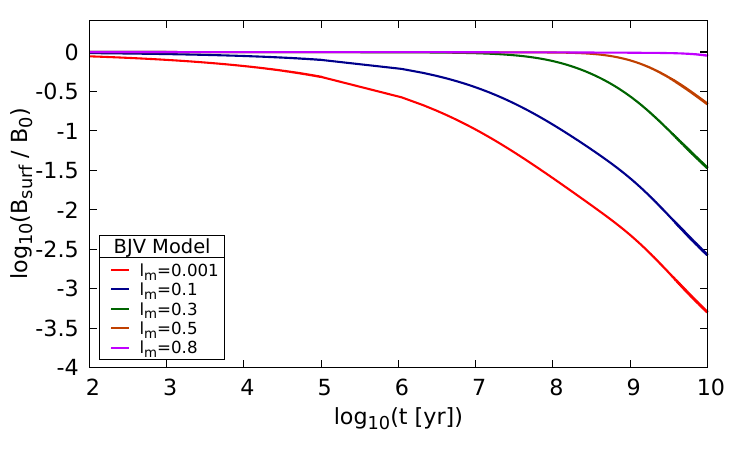} \\
    \includegraphics[width=\columnwidth]{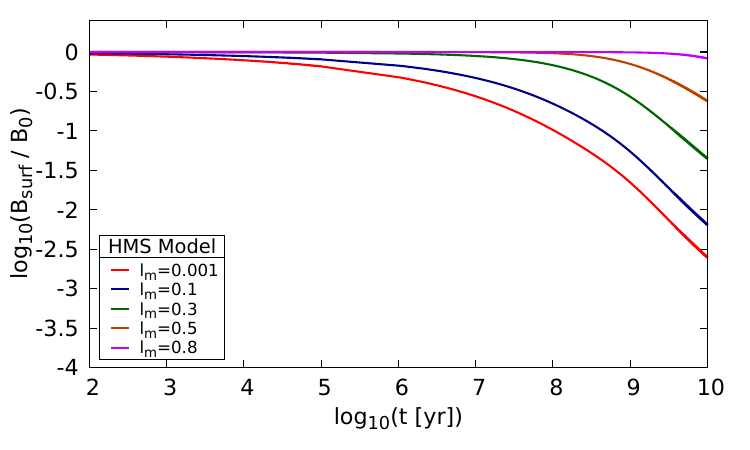}
    \caption{Normalised magnetic field decay for the Walecka (top), Bethe-Johnson V (middle), and HMS (bottom) EoS for an impurity parameter value of $Q=0.001$.}
    \label{fig:fan_plot}
\end{figure}

The temporal evolution of these fields over $10^{10}$ years is shown in Figure~\ref{fig:fan_plot}. Here, we plot the normalised surface magnetic field decay for all three models for a fixed impurity parameter value of $Q=0.001$. The results demonstrate a ``fanning out'' effect, clearly indicating the depth influence on the decay timescales.

For shallow fields ($\lm \lesssim 0.1$), the magnetic field decays by approximately 3 orders of magnitude in $10^{10}$ years, easily accommodating the 1--2 orders of magnitude decay observationally required for older pulsar populations. Conversely, for fields anchored deeper in the crust ($\lm \gtrsim 0.5$), the higher conductivity of the inner crust effectively `freezes' the magnetic flux in place. These deep fields show negligible decay (up to $\sim 0.5$ orders of magnitude) over the entire $10^{10}$ year period, reinforcing the constraint that pulsars cannot host deeply anchored fields if they undergo standard Ohmic decay.

\begin{figure}
    \centering
    \includegraphics[width=\columnwidth]{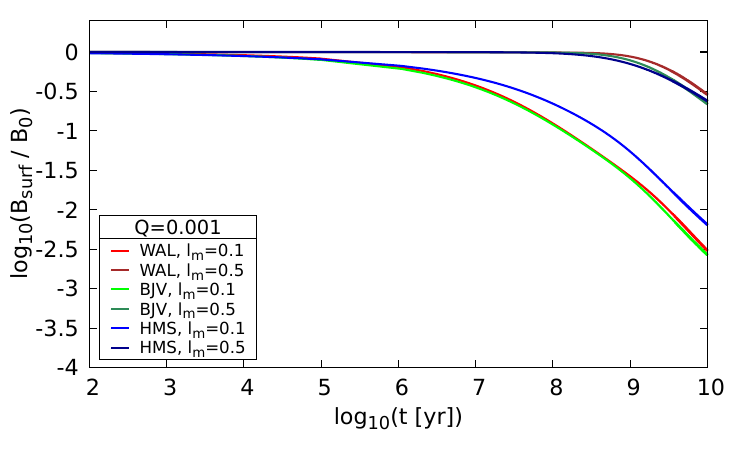}
    \caption{Comparison of decay rates between the Walecka, Bethe-Johnson V, and HMS models for two values of $\lm = [0.1, 0.5]$}
    \label{fig:eos_compare}
\end{figure}

To ensure these decay trends are not artefacts of a specific stellar model, we directly compare the $\lm = 0.1$ and $\lm = 0.5$ configurations across all three EoS models in Figure~\ref{fig:eos_compare}. 

Despite the significant variance in physical crust thickness ($\Delta R \approx 1.33$ km for Walecka vs $\Delta R \approx 0.67$ km for the HMS model), the decay trends are qualitatively identical. We note that the more compact HMS model exhibits a marginally slower decay rate compared to $1.4 \, \msun$ models (WAL and BJV). Nonetheless, the normalised depth parameter $\lm$ remains the overwhelmingly dominant factor in dictating the field's lifespan.

\begin{figure}
    \centering
    \includegraphics[width=\columnwidth]{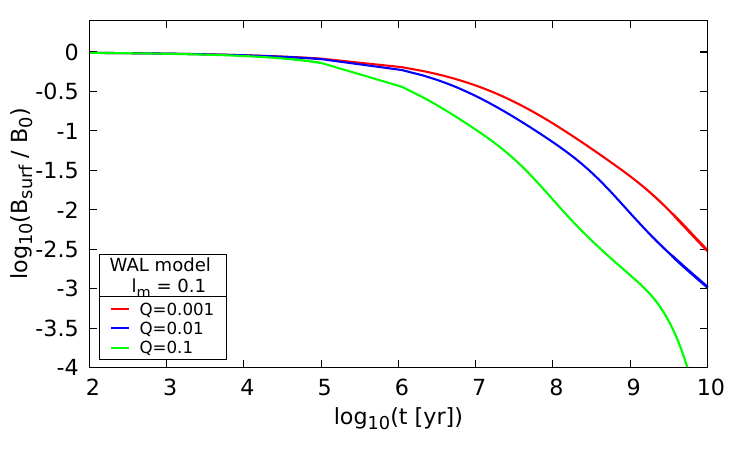}
    \caption{The effect of the impurity parameter $Q$ on an intermediate field depth ($\lm=0.1$) using the Walecka model EoS.}
    \label{fig:impurity}
\end{figure}

We then investigate the role of crustal impurities in Figure~\ref{fig:impurity}. We demonstrate the effect of the impurity parameter on the decay by looking at one specific set of parameters, that being for an intermediately anchored field ($\lm = 0.1$) using the Walecka EoS. It is clear that increasing the impurity parameter value from $Q=0.001$ to $Q=0.1$ accelerates the decay rate of the surface magnetic field. By $10^{10}$ years, the more impure crust permits the field to decay by more than 4 orders of magnitude, compared to only 2.5 orders of magnitude in the purer crust. Thus, while $\lm$ predominantly controls the decay rate of the field, the impurity parameter $Q$ does also have an effect.
 
Because lower $Q$ values yield higher electrical conductivity, Ohmic dissipation is significantly slowed. Therefore, to reproduce the decay trends observed in older pulsar populations under highly conductive conditions, the magnetic field should be pushed outward, concentrating the currents very near the stellar surface where resistivity is highest. Conversely, a crust with a larger impurity parameter ($Q=0.1$) with accelerated Ohmic decay would allow a field to remain anchored slightly deeper while still matching the same observational trends. However, the high-density inner crust is simply too conductive to support significant decay regardless of outer-crustal impurities. These results therefore indicate that significant field evolution requires the magnetic flux to be anchored primarily in the shallow outer crust, even under the lowest-conductivity assumptions.

\begin{figure}
    \centering
    \includegraphics[width=\columnwidth]{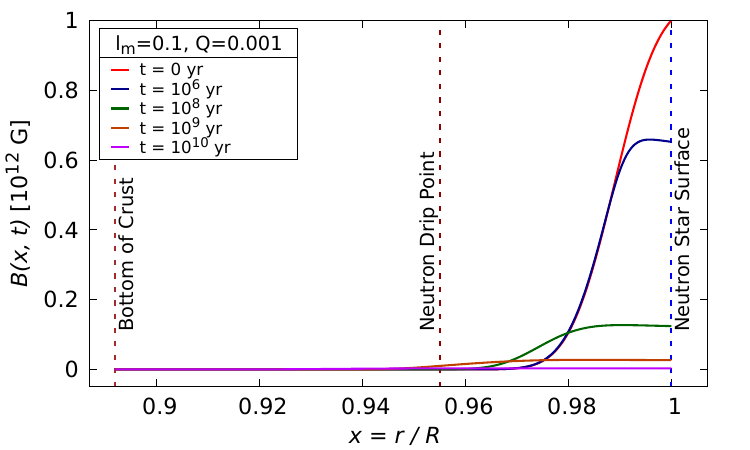} \\
    \includegraphics[width=\columnwidth]{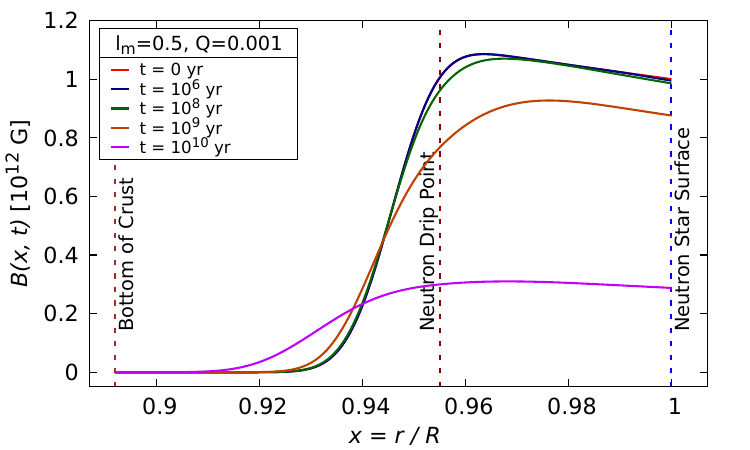}
    \caption{Radial magnetic field evolution for the Walecka model with $Q=0.001$, shown for $\lm=0.1$ (top) and $\lm=0.5$ (bottom) at selected ages.}
    \label{fig:radial_evolution}
\end{figure}

We have also examined the magnetic field profiles $B_r(x,t)$ for representative shallow and deeper configurations in the Walecka model with $Q=0.001$. Figure~\ref{fig:radial_evolution} shows the radial distribution of the magnetic field throughout the crust at selected evolutionary times, $t=0$, $10^6$, $10^8$, $10^9$, and $10^{10}$ yr, for $\lm=0.1$ and $\lm=0.5$. For the shallower $\lm=0.1$ configuration, the radial profile changes appreciably as the field decays, with the surviving field becoming progressively weighted towards deeper crustal layers. In contrast, the $\lm=0.5$ profile remains nearly unchanged over the earlier evolution and develops noticeable redistribution only at later times, when its field strength also begins to decay.

\section{Discussion}
\label{sec:Disc}
In this work, we have considered the long-term evolution of a crust-confined, axisymmetric dipolar magnetic field in isolated radio pulsars, isolating the contribution of Ohmic diffusion. We adopt an isothermal crust and neglect Hall drift, while exploring the effects of initial field depth, crustal impurity, and stellar structure across three different EoS models. This framework allows us to isolate how the radial location of the magnetic field and the corresponding conductivity profile influence the long-term Ohmic decay timescale.

Our results demonstrate that the decay timescales of crustal magnetic fields are largely governed by their radial depth, $\lm$. Our magnetic field evolution results, presented in terms of the normalised field $B(t)/B_0$, establish predictions for the fractional decay magnitude that are applicable to standard isolated radio pulsars ($B \sim 10^{10}$ -- $10^{12}$ G). These models are specifically tailored to the physical conditions of this broadly observed population, where Ohmic dissipation drives the late-time evolution. For standard radio pulsars, which are expected to remain active for $\sim 10^7$--$10^8$ yr before crossing the pulsar death line, our results (see Fig.~\ref{fig:fan_plot}) indicate that the magnetic field cannot extend deeply into the inner crust if any significant observational decay is to occur over their lifetimes.

The radial profiles in Figure~\ref{fig:radial_evolution} also show that $\lm$ is better interpreted as an initial depth parameter rather than a permanently fixed anchoring radius: configurations that undergo appreciable decay can also redistribute within the crust, whereas deeper fields remain comparatively stable until much later times. This behaviour, however, does not alter the dominant depth dependence identified above.

The adopted temperature of $T=10^{5.5}$ K in our work also provides a
conservative thermal assumption in the context of this result. Within our
conductivity prescription, a lower crustal temperature reduces electron-phonon
scattering and therefore increases the electrical conductivity, further
suppressing Ohmic dissipation. Consequently, our results imply that adopting a lower temperature would require the initial current distribution to be
confined to still shallower crustal layers in order to reproduce comparable
observational decay rates.

Furthermore, we observe a distinct dependence on the underlying stellar structure. In our analysis, we compared models spanning a range of physical crustal thicknesses: the extended crust of the Walecka model ($\Delta R \approx 1.33$ km), the intermediate Bethe-Johnson V model ($\Delta R \approx 0.98$ km), and the highly compressed higher mass star model ($\Delta R \approx 0.67$ km). Despite these stark differences in absolute geometric thickness, all three models yield comparable decay trends when normalised by fractional depth. However, we note that the higher mass configuration exhibits a marginally slower decay rate for an equivalent fractional depth compared to the canonical $1.4 M_{\odot}$ models (WAL and BJV). This suggests that more massive neutron stars tend to be more efficient at retaining their magnetic fields. Consequently, a massive neutron star could inherently appear to possess a more stable magnetic field than a lower-mass counterpart.

Finally, our study supports the premise that the magnetic fields of isolated neutron stars are anchored predominantly near the surface, within the outer crust, to optimally reconcile theoretical diffusion models with observational data. Although individual spin-down ages provide uncertain estimates of the true chronological ages of pulsars, population synthesis studies provide broader constraints on the long-term magnetic field evolution through the observed $P-\dot{P}$ distribution. Within our conductivity prescription, the high conductivity of the deep inner crust suppresses significant Ohmic dissipation on astrophysically relevant timescales. Reproducing the population-level magnetic field evolution inferred for radio pulsars therefore favours supporting current distributions confined to the more resistive outer layers of the stellar crust. This work provides a simplified treatment within the broader study of neutron star magnetic field evolution. This study can be further enhanced to include additional physical effects, including thermal evolution, Hall drift, and more general magnetic field configurations in future investigations.

\section*{Acknowledgements}
KS acknowledges support from the National Research Foundation of South Africa (NRF). 

\section*{Data Availability}
No new data were generated. The code used to generate the models will be made available upon reasonable request.

\bibliographystyle{mnras}
\bibliography{refs}

\label{lastpage}
\end{document}